\pdfoutput=1
\documentclass[aps,prx,reprint,floatfix,superscriptaddress,showpacs,longbibliography]{revtex4-1}
\usepackage[T1]{fontenc}

\usepackage{color}

\usepackage{braket}
\usepackage{amsmath}
\usepackage{amssymb}
\usepackage{csquotes}
\usepackage{graphicx}
\usepackage{epstopdf}
\usepackage{tikz}
\usetikzlibrary{calc}

\usepackage{xcolor}
\usepackage{hyperref}
\usepackage{bookmark}
\usepackage{array}
\usepackage[normalem]{ulem}

\newcommand*{\floor}[1]{\lfloor #1 \rfloor}
\newcommand*{\gsh}{g_{\textrm{\tiny Sh}}}

\begin{document}




\title{Signature of the Chiral Anomaly in Ballistic Magneto-Transport}

\author{Thibaud Louvet}
\affiliation{Universit\'e de Lyon, ENS de Lyon, Universit\'e Claude Bernard, CNRS, Laboratoire de Physique, F-69342 Lyon, France}

\author{Manuel Houzet}
\affiliation{Univ.~Grenoble Alpes, CEA, INAC-Pheliqs, F-38000 Grenoble, France}

\author{David Carpentier}
\affiliation{Universit\'e de Lyon, ENS de Lyon, Universit\'e Claude Bernard, CNRS, Laboratoire de Physique, F-69342 Lyon, France}

\date{\today}

\begin{abstract}
We compute the magneto-conductance of a short junction made out of a Weyl semi-metal.  We show that it displays quantum oscillations at low magnetic field and low temperature,
before reaching a universal high-field regime where it increases linearly with the field. 
Besides identifying a new characterization of the physics of materials with 
Weyl singularities in their energy spectrum, this ballistic regime corresponds to the 
simplest setup to study and understand the manifestation of the so-called chiral anomaly of massless relativistic particles in condensed matter.  
At low fields the  algebraic in field magneto-conductance  incorporates contributions besides the anomalous 
chiral current while the linear conductance at higher fields 
constitutes an unambiguous signature of the chiral anomaly in a Weyl conductor. 
Finally, we study the dependence of the ballistic magneto-conductance on the chemical potential, and
discuss the cross-over towards the diffusive regime when elastic scattering is present.
\end{abstract}


\maketitle


\section{Introduction}

When relativistic particles in three dimensions become massless, they acquire an additional symmetry: chirality. However, while this chirality is a conserved quantity of the Hamiltonian, it is no longer conserved by the associated field theory. This very intriguing and unanticipated property was called a chiral anomaly, or
  Adler-Bell-Jackiw anomaly~\cite{Adler:1969,Bell:1969}. 
  Much efforts have been devoted to identifying measurable consequences of this anomaly. In particular, it was 
  realized that, in the presence of a magnetic field,  a difference of chemical potential  between particles with opposite chiralities induces a chiral current: the chiral magnetic effect \cite{Kharzeev:2014}. 

In recent years, the prospect of probing consequences of the chiral anomaly, which was initially discussed 
in the context of high energy physics, through the transport properties of solids has brought the study 
of this effect in the realm of condensed matter physics. 
 Indeed, in several materials electrons behave as relativistic particles.  Examples include Weyl and Dirac semi-metals, in which 
the Fermi energy lies close to a linear crossing between energy bands. The study 
of such crossing is not new \cite{Herring:1937}. However, it has been completely revived by the study of topological properties 
associated with these band crossings in the fields of Helium physics \cite{Volovik:2003}, strongly correlated iridates 
\cite{Wan:2011},
and, more generally, the topological band theory of materials~\cite{Bansil:2016}.  
 The search for manifestations of the chiral anomaly 
 was associated with the prediction of 
 the positive magneto-conductance in materials with linear band crossing. 
Indeed, the presence of both electric ${\bf E}$ and magnetic ${\bf B}$ fields induces a chiral current proportional to ${\bf E}.{\bf B}$ and modifies the associated electromagnetic response of the massless relativistic particles. 
 This chiral anomaly was discussed in the context of solids by Nielsen and Ninomiya  who considered a lattice realization of a Weyl semi-metal~\cite{Nielsen:1983} (see also Ref.~\onlinecite{Aji:2012}). 
They related a positive magneto-conductance along ${\bf B}$ in the quantum regime to the anomalous 
 Landau level in the presence of inter-cone scattering processes.
 Recently, the anomalous response of Weyl particles to weak electromagnetic fields, which was analyzed by taking into account Berry curvature corrections in the semiclassical equations of motion, was also attributed to the chiral anomaly \cite{Son:2013,Burkov:2014}.
This recent progress is reviewed in Ref.~\cite{Burkov:2015}. 
 The above cited works characterized the linear response of bulk electrons submitted to both 
electric and magnetic fields. Alternatively, recent proposals considered the so-called chiral magnetic effect induced by an oscillating magnetic field or chiral chemical potential in a junction~\cite{Chen:2013,Zhong:2016,Baireuther:2015}. 

On the experimental side, the magneto-conductance was studied in various materials identified as candidates 
for this relativistic physics, including 
Cd$_3$As$_2$ \cite{Li:2015}, 
Na$_3$Bi  \cite{Xiong:2015}, 
TaAs \cite{Huang:2015}, 
NbP \cite{Niemann:2017}. 
In these compounds, the positive magneto-conductance for a parallel magnetic field, as well as its extreme 
sensitivity to the orientation of the magnetic field with respect to the electric field, were taken as manifestations of the 
underlying chiral anomaly of the relativistic massless equations of motion. 
 More recently, the prospect of probing the chiral anomaly in the so-called hydrodynamic regime, in which elastic 
 scattering of electrons is dominated by the Coulomb interaction, has led to an experimental study of the 
 magneto-conductance of WP$_2$~\cite{Gooth:2017}. 
In this material, an elastic mean free path of the order of $100\mu m$ at 4K, that is, larger than the sample size, was detected. 
 While scattering was claimed to be dominated by interactions between electrons, lowering even further the temperature in such a sample would drive the conductor into the so-called ballistic regime. 
 
The purpose of the present work is to study the longitudinal magneto-conductance of ballistic 
junctions made with Weyl semi-metals in the regime where equilibration takes place only in the leads. 
This corresponds to so-called 
cold electrons, as opposed to hot electrons when energy relaxation occurs within the conductor. 
 We find that two regimes must be distinguish : (i) a regime in which the conductance behaves algebraically 
 at low fields and displays quantum oscillations at higher fields
 (ii) a quantum regime reached for magnetic fields such that  
 $B \lambda^2_F \geq h/e$ where $\lambda_F$ is the Fermi wavelength, 
 in which the conductance is linear in field with a universal slope. 
 Owing to the conceptual simplicity of the ballistic  regime 
 we argue that only the magneto-conductance in the quantum regime is an unambiguous signature of 
 the chiral anomaly, and can be experimentally accessed for a chemical potential close enough to 
 the band crossing.

In Sec.~\ref{sec:2}, we determine analytically the
ballistic magneto-conductance at zero temperature for the simplest model of two well-separated 
(in momentum space) Weyl cones with opposite chiralities. We find that the magneto-conductance displays quantum
oscillations as the magnetic field increases, similar to Shubnikov-de Haas oscillations, but along the direction of
 the field.  These quantum oscillations evolve into a robust linear 
 magneto-conductance at large field. On the other hand, we discuss how the low-field oscillations are suppressed 
 by the temperature, or in the presence of a broadening of the Landau levels due to their
coupling with the leads. In Sec.~\ref{sec:3}, we compare our predictions with the numerics for a tight-binding 
model with four pairs of Weyl cones. We obtain a good agreement with the analytics at low chemical potential, 
when Weyl cones are well separated in momentum space. On the other hand, we find that the slope of the linear
 magneto-conductance at large magnetic field decreases by a factor two as the chemical potential increases above 
the saddle-point energy corresponding to the merging of two Weyl points with opposite chirality. We explain this 
effect as the signature of the assymmetry of the magneto-conductance with the magnetic field for a single 
pair of Weyl cones,
as the chemical potential is tuned away from the band crossing. 
The symmetry is actually recovered in the specific model with eight cones, due to their distribution in momentum space. We confirm this interpretation with numerics
for a tight-binding model with a single pair of Weyl cones.  In Sec.~\ref{sec:4}, we analyze the 
magneto-conductance within a ballistic semiclassical theory that allows recovering its linear-in-field dependence 
in the quantum regime. The quantum oscillations are beyond the semiclassical approximation, and we obtain 
a quadratic magneto-conductance at small magnetic field, which is distinct from the predictions of 
Sec.~\ref{sec:2} in the presence of level-broadening at low field. Furthermore, we study the stability of the 
linear magneto-conductance in the presence of elastic intra-cone and inter-cone scattering, 
we find that it is pushed toward very large fields in strongly disordered systems, in agreement with 
Ref.~\cite{Altland:2016}. In Sec.~\ref{sec:5}, 
we discuss our results in the context of the 
chiral anomaly, and we argue that only the quantum regime is a clear signature of it, before concluding 
in Sec.~\ref{sec:6}.

\section{Ballistic Magneto-Transport}
\label{sec:2}
\begin{figure}[tb]
\begin{center}
\includegraphics{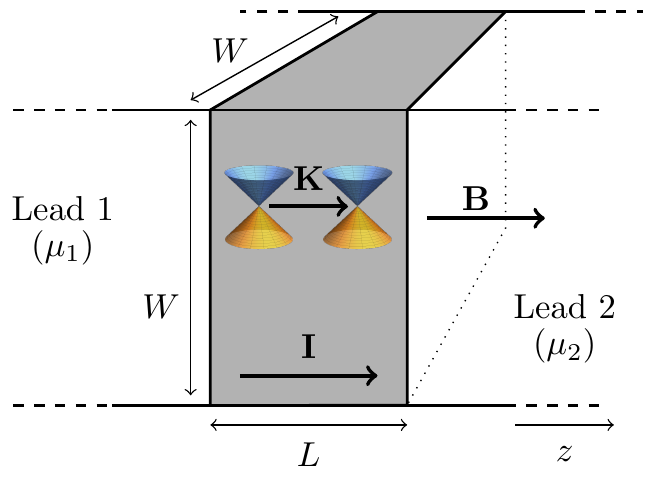}
\caption{
\label{fig:DessinJunction} 
      Schematic representation of a short junction of a Weyl semi-metal between two metallic leads
      at chemical potentials $\mu_1,\mu_2$, and submitted to a magnetic field ${\mathbf B}$ parallel to the direction
      of the current ${\mathbf I}$. 
      ${\mathbf K}$ is the vector in the Brillouin zone relating a pair of Weyl cones of opposite chiralities. 
      }
\end{center}
\end{figure}
We consider the transport through a short junction made of a Weyl semi-metal in the presence of an external magnetic field ${\bf B}$, represented in Fig.~\ref{fig:DessinJunction}. 
We assume that ${\bf B}$ is applied along the direction of the current, unless specified otherwise.
In such a setup, the charge current is induced by the bias voltage between the leads on 
each side of the junction. For a short enough junction, scattering inside the conductor can be neglected.
Hence, the energy and momentum relaxation only take place in the leads.  
This so-called ballistic regime is inherently out of equilibrium, in contrast with the linear response to an electric field, which has been considered so far. 
The current induced by a small difference of chemical potentials across the junction, 
$\mu_{2/1} = \mu \mp \delta \mu/2$, 
takes the simple form 
$ I =  G  (\delta \mu/e)$. The conductance of the junction, $G$, scales as its transverse area, $W\times W$, 
and does not depend on the junction's length, $L$, in the ballistic regime. This allows us defining a dimensionless and 
scale independent conductance, $g$, such that 
\begin{equation}
	G = \frac{e^2}{h} \left( \frac{W}{a}\right)^2 g, 
\end{equation}
where $a$ is an UV cutoff (for instance, the lattice spacing). 

\begin{figure}[htbp]
\begin{center}
	\includegraphics[width=8.5cm]{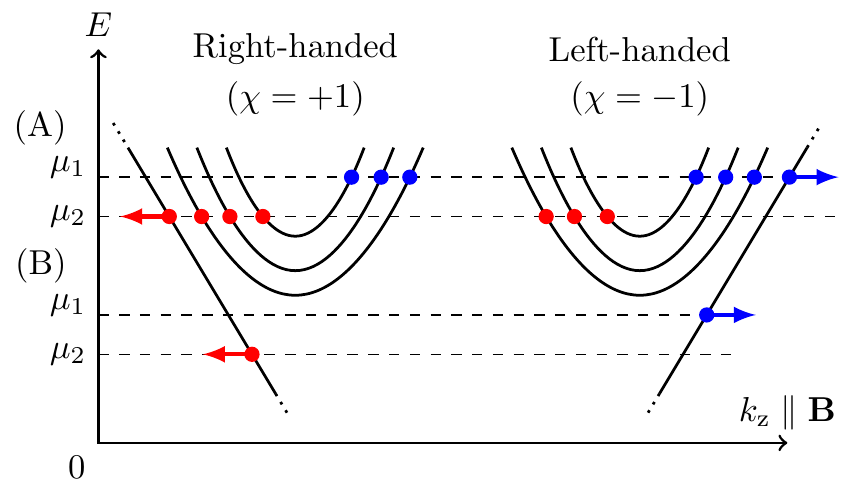}
  	\caption{\label{fig:LandauLR} 
  	Schematic view of the Landau level dispersion for a pair of left and right handed Weyl cones.
   	All left movers (red) come from the lead $2$ at chemical potential 
  	$\mu_2$ and all right-movers (blue) come from the lead $1$ at $\mu_1$. 
  	An equilibrium chiral current crosses the junction irrespective of the chemical potential, due to the presence of
  	the $n=0$ chiral Landau level.
  	The out-of-equilibrium charge current is induced by the difference $\mu_1 - \mu_2$. The corresponding ballistic 
  	conductance is proportional to the number of Landau levels intercepted by the chemical potential (case A). The quantum regime is reached when only 
  	the level $n=0$ crosses the chemical potential (case B). 
  	}
\end{center}
\end{figure}

In the presence of a magnetic field ${\bf B} \parallel z$, the kinetic energy of electrons in the 
$x,y$-directions 
freezes into Landau levels, while the motion in the $z$-direction is unaffected. Hence each state in each 
Landau level provides a conduction channel in the ballistic junction. 
 The number of such channels per Landau level is $W^2 / (2\pi l_{B}^{2})$ where $l_{B}=\sqrt{\hbar/(e|B|)}$ is the
magnetic length. This yields a dimensionless conductance 
\begin{equation}
	g(\mu,b) = N(\mu,b) ~|b|,
\label{eq:g_b}
\end{equation}
where $N(\mu,b)$ is the number of Landau levels below the chemical potential $\mu$ and $b$ is the rescaled magnetic flux per unit cell (in units of the flux quantum), that is,  
$
b =  \phi/\phi_{0} =  a^{2}B e/ h = \text{sign}(B) ~ a^{2} / (2\pi l_{B}^{2}),
$

Let us now study the evolution of $N(\mu,b)$ for a ballistic junction built out of Weyl fermions. Weyl valleys necessarily 
come by pairs of chiral fermions with opposite chiralities. Around each valley, the local Bloch Hamiltonian can be written  
as 
\begin{equation}
\label{eq:model}
H = 
c p_x \sigma_x + c p_y \sigma_y + 
\chi c_z p_z \sigma_z.
\end{equation}
with $\chi = \pm 1$ the right/left-handed chirality of the  valley, 
$p_{i} = \hbar k_{i}$ are quasimomenta, while $c$ and $c_z $ are velocities in $x,y$- and $z$-directions 
respectively. (For simplicity, we assumed same velocities along $x$ and $y$.)
 In the presence of a magnetic field ${\bf B} \parallel z$, the spectrum of each valley consists of a series of bands 
$E_{n,\pm} = \pm [ n   (\hbar \omega_0)^2 +  \left( c_z \hbar k_z \right)^2 ]^{1/2}$ with
$n\geq 1$, which disperse along the $z$-direction, and which are
separated from each other by gaps at $k_z=0$ of the order 
 $\hbar \omega_0 = \hbar c  \sqrt{2} /  l_{B} = \epsilon_0 \sqrt{|b|}$ with
$\epsilon_0 = \hbar c \sqrt{4 \pi}/a$ (see Sec.~I of the supplementary material (SM) \cite{SupplMat}). 
Besides these bands, the energy spectrum admits an additional, linearly dispersing band, whose direction of propagation 
 depends both on the chirality of the valley and the direction of the magnetic field,
$E_{0}  = - \chi ~ \textrm{sgn}(B) ~ c_z \hbar k_z$, as shown in Fig.~\ref{fig:LandauLR}. 
\begin{figure}[htbp]
\begin{center}
	\includegraphics[width=8.5cm]{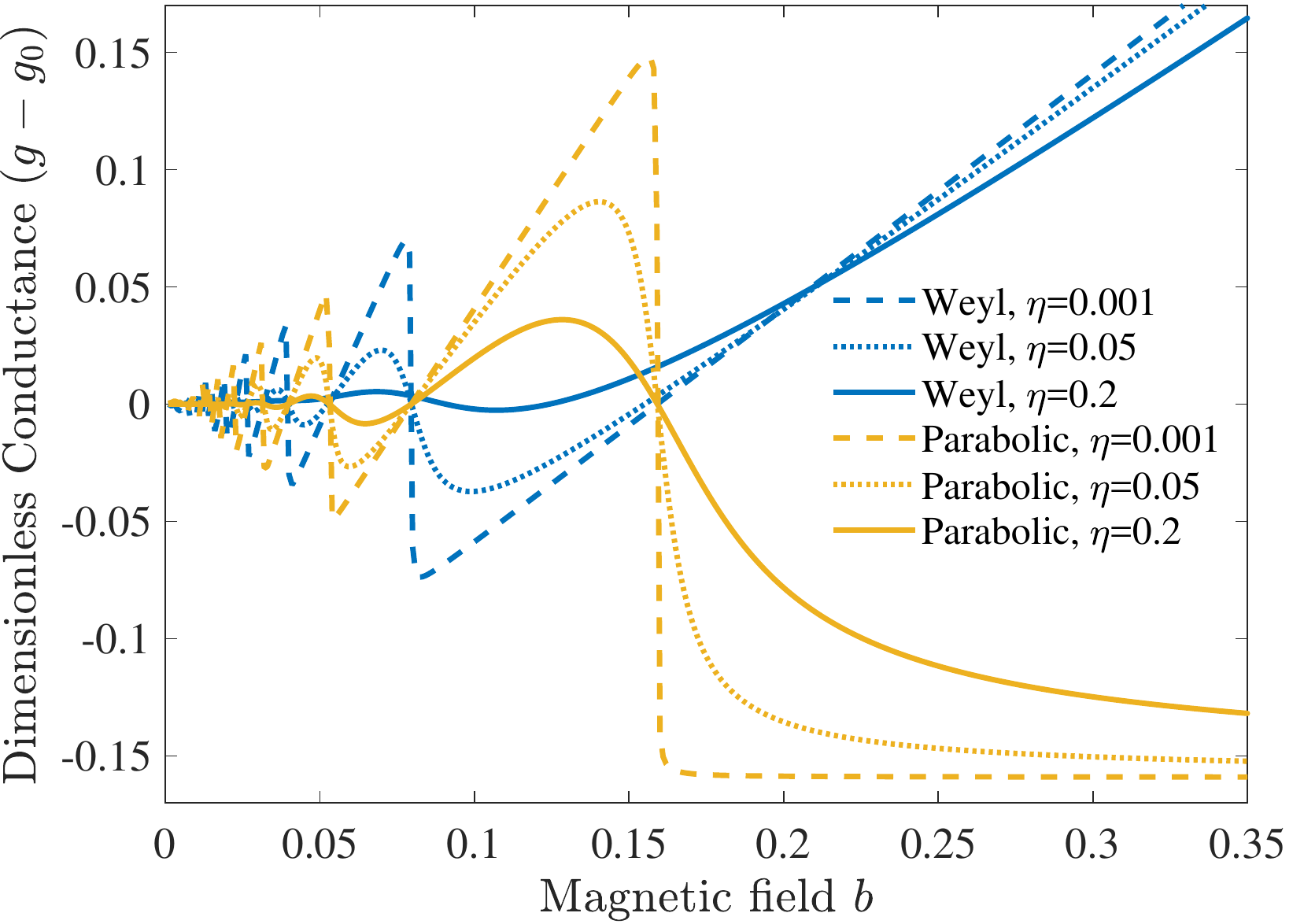}
  	\caption{\label{fig:Theory} Magneto-conductance of ballistic junctions with Weyl (blue) and parabolic (yellow)
  	 dispersion relations. 
  	 The finite dwell time of the electrons in the junction is included through the broadening 
  	 amplitude $\eta$ of the spectrum and we chose $\mu=0.7,c=1$. 
   	 For small magnetic fields, several Landau bands contribute to the conductance. Correspondingly the 
   	 conductance oscillates as a function of the magnetic field, reflecting the discontinuous change of the number of filled Landau bands as the later varies.
  	 At high magnetic fields, in the quantum regime, 
  	 the Weyl and parabolic magneto-conductances differ. In this regime a single, $n=0$ Landau 
  	 band contributes to the conductance of the Weyl junction, and it increases linearly with $|b|$. By contrast,
  	 for a junction with a parabolic dispersion relation, no Landau band crosses the chemical potential, and the 
  	 conductance vanishes. 
  	  	}
\end{center}
\end{figure}
At large magnetic field, $|b|\geq \mu/\epsilon_0 $, only the anomalous Landau level $n=0$ contributes 
to transport, a situation reminiscent to that of Ref.~\cite{Nielsen:1983} albeit considered here in the ballistic 
regime.
We call this regime the quantum regime, in which $N(\mu,b)=1$, and from Eq.~\eqref{eq:g_b} we find
$ g = |b| $. 
At smaller fields, $|b|\leq \mu/\epsilon_0 $, 
other Landau bands intercept the Fermi level and contribute to the conductance, which thus oscillates as a function of the magnetic field $|b|$ or the chemical potential $\mu$.
For an ideal ballistic junction, the number of filled Landau levels is 
deduced from the expression of Landau levels, 
$	
N(\mu,b) = 1 + 2 \floor{ (\mu/(\hbar \omega_0))^2 }
	 = 1 + 2 \floor{  \mu^2 / ( \epsilon_0^2 |b| ) },
$
 where $ \floor{x} $ stands for the integer part of $x$. 
By introducing the sawtooth function $\textrm{sw}(x) = \floor{x} - x + \frac12 $ we obtain the expression of the 
conductance of an ideal Weyl junction, 
\begin{equation}
	g(\mu,b) = \gsh(\mu) + 2|b| ~ \textrm{sw}\left(\frac{\gsh (\mu)}{2|b|} \right),  
\label{eq:g_ideal}
\end{equation}
where 
$\gsh (\mu)
= 2 \mu^2 / \epsilon_0^2 = a^2 \mu^2/(2\pi (\hbar c)^2)  $  
corresponds to the Sharvin conductance of the junction, which is determined by the number of 
conduction channels at $b=0$.
Let us note that quantum oscillations of the longitudinal 
conductance, similar to those we identified in the ballistic regime, 
were also observed in the different  regime of diffusion of hot electrons close to equilibrium 
within a Kubo formula approach \cite{Gorbar:2014}.

%
 \begin{figure*}[!thb]
\centerline{
\includegraphics[height=7.5cm]{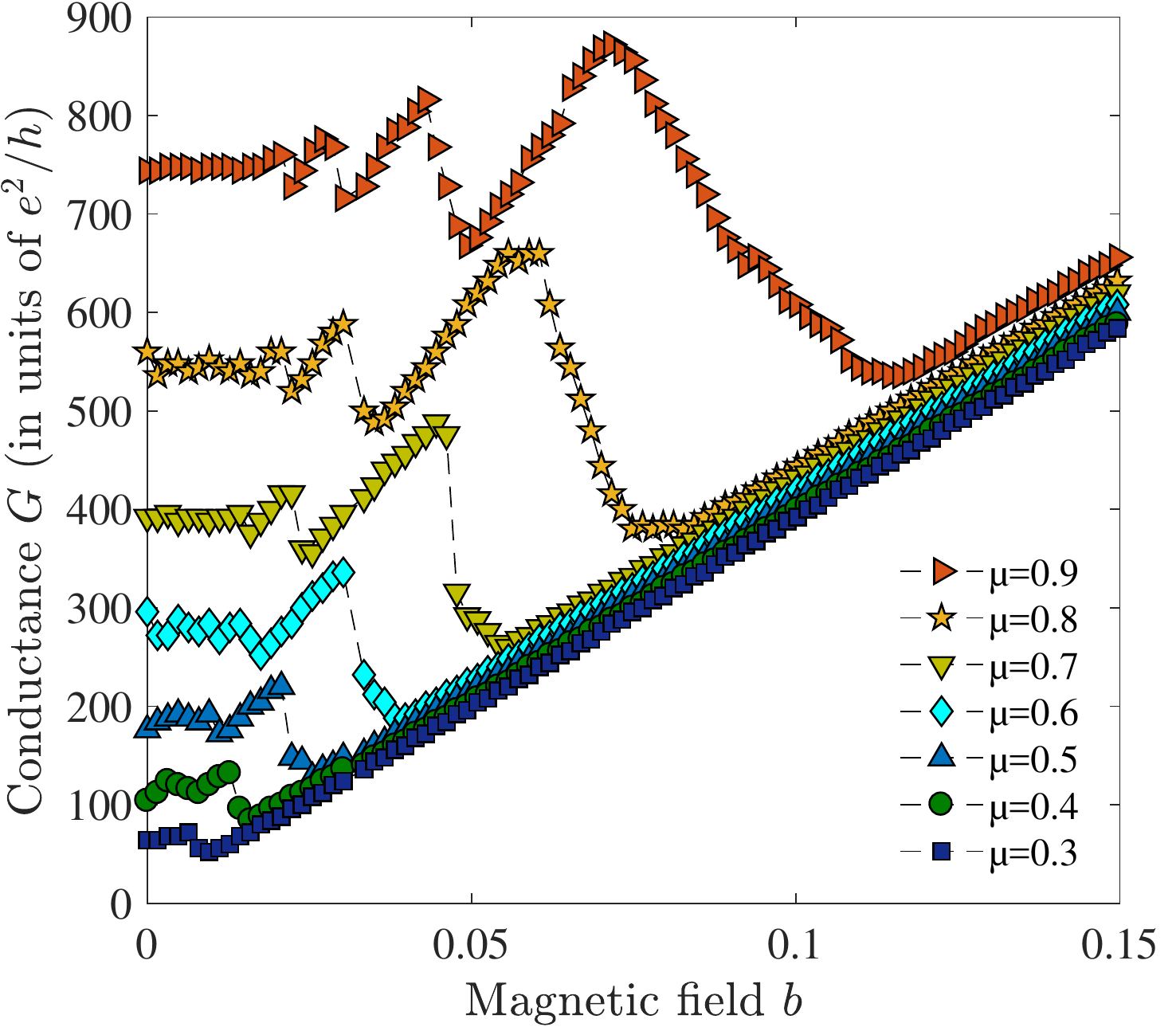}
\includegraphics[height=7.5cm]{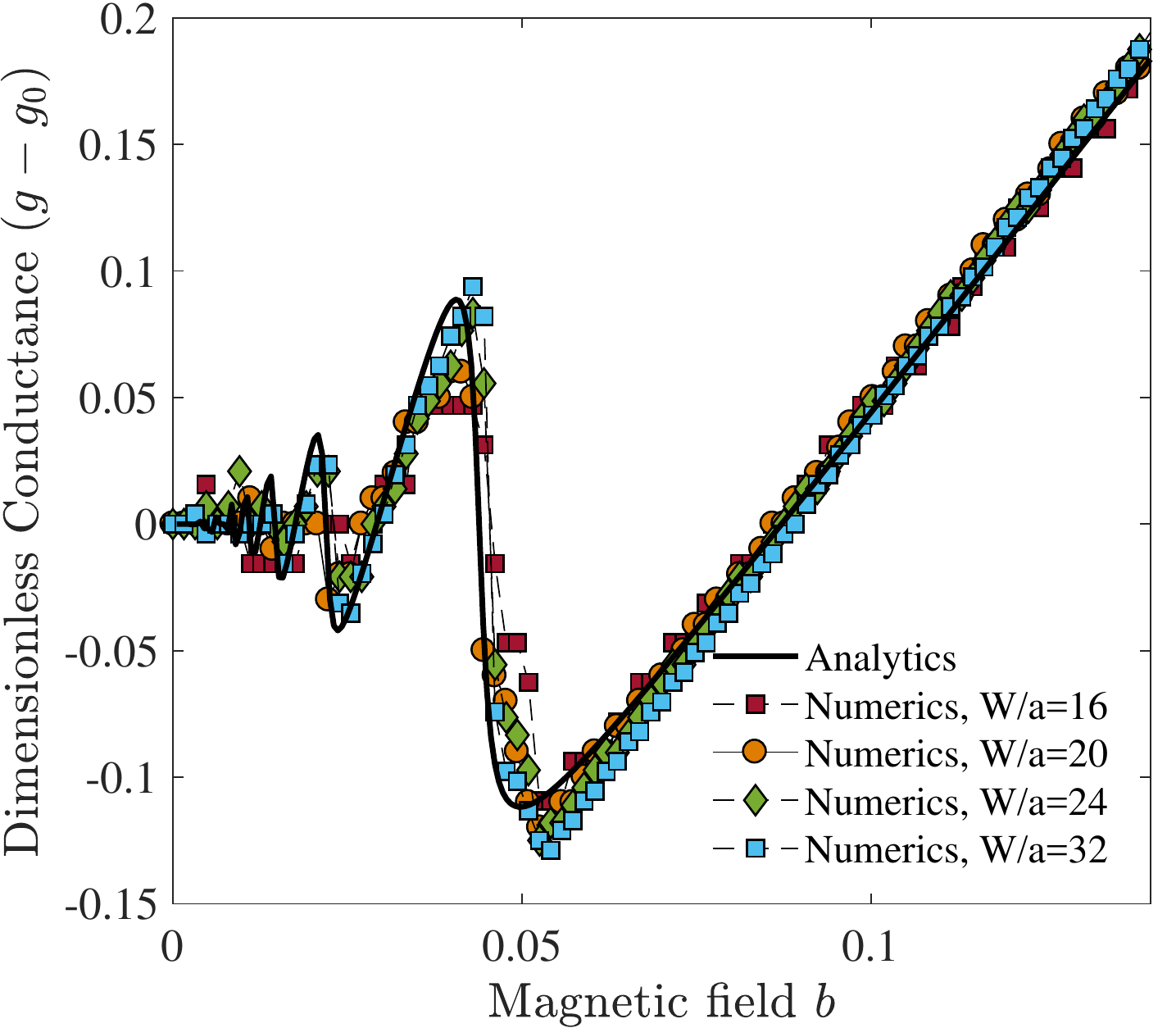}
}
	\caption{\label{fig_numericalConctance}  
Conductance of a short ballistic Weyl junction, in units of the conductance quantum, as a function of the magnetic field along the junction, in units of the flux per lattice unit cell. At low magnetic field, the conductance oscillates before 
	reaching a linear regime at high field, with a universal and positive slope. 
	A lattice model with 8 Weyl cones is used. 
	The left panel shows the magneto-conductance for a junction of transverse size $W/a=32$, for various chemical potentials.
	The right panel shows the collapse of rescaled dimensionless conductances $g$ 	for a fixed $\mu=0.7$, and comparison (full line) with the expression \eqref{eq:g-broaden} incorporating 
	a Lorentz broadening of energy levels due to the leads ($\eta=0.01$). }
\end{figure*}

The limit $b\to 0$ of the conductance \eqref{eq:g_ideal} for an ideal junction is actually ill-defined: 
as usual, we expect the quantum oscillations described by Eq.~\eqref{eq:g_ideal} to be cut-off at low field 
by any broadening of the Landau bands 
\cite{Ashby:2014}. In the ballistic junction that we consider, the finite dwell time of the electrons between the 
leads, $\tau_d = L/c_z$ provides an inherent broadening of order 
$\eta \simeq \hbar / \tau_d =\hbar c_z / L$. 
We can phenomenologically incorporate this effect by a standard Lorentz broadening of the energy levels, 
leading to~\footnote{Note that the Lorentz broadening of the Sharvin conductance, $g(\mu,b=0)$, requires a high-energy 
cutoff below which the Hamiltonian \eqref{eq:model} is defined, as opposed to the $b$ dependent correction described by Eq.~\eqref{eq:g-broaden}.}
\begin{multline}
	g(\mu,b)- g(\mu,0) 	= \\
	2 |b|
    \int  \frac{d\mu'}{\eta} f \left(\frac{\mu-\mu'}{\eta} \right)   
     \textrm{sw}\left(\frac{\gsh(\mu')}{2|b|} \right), 
\label{eq:g-broaden}
\end{multline}
with $f(x) = 1/[\pi (1+x^2)]$.
In the regime where a non-oscillating density of states is recovered, corresponding to 
$\hbar \omega_0 = \epsilon_0 |b| \ll  \eta $,  we find a non-analytic scaling of the positive magneto-conductance,
\begin{equation}
 g(\mu,b)- g(\mu,0) \simeq \alpha~  \frac{\epsilon_0}{\eta} f\left(\frac{\mu}{\eta}\right) ~  b^{3/2}
  \label{eq:largebroad}
\end{equation}
with $\alpha \simeq 1.16$, see Sec.~II in the SM~\cite{SupplMat}.
Noteworthy, this scaling is different from the typical $b^2$-behavior predicted in the diffusive regime~\cite{Son:2013}. 
Similarly if a  thermal broadening of the levels supersedes the intrinsic broadening, we can still use 
Eq.~\eqref{eq:g-broaden}, but now with $f(x)=-f'_F(x)$ with $f_F (x) = 1/(1 + e^x)$.
For Weyl fermions, it yields 
\begin{multline}
g (\mu,b) - g(\mu,0) = \\
	|b| ~ f_F\left(\frac{\mu}{k_B T}\right) 
	- |b|^{\frac32} \frac{\epsilon_0}{4 k_BT}  f_F'\left(\frac{\mu}{k_B T}\right).
\label{eq:euler-highT}
\end{multline}
%
%
In particular, the anomalous magneto-conductance is exponentially suppressed at $\mu\gg k_BT$.

Let us contrast this behavior with that for a standard non-relativistic parabolic dispersion relation 
$E = p^2 / (2m)$, whose Landau bands read
	$ E_{n} =  \hbar \tilde{\omega}_0 (n-\frac12 ) + p_z^2/(2m)$  with $n\geq 1$, 
	$ \hbar \tilde{\omega}_0 = \hbar e |B|/m = \epsilon_1 |b| $,
and 
	$\epsilon_1 = h^2/(2\pi m a^2) $. 
The number of filled Landau bands is now 
	$N(\mu,b) =  \floor{ \mu/(\hbar \tilde{\omega}_0) + \frac12 } $ 
corresponding to a conductance 
$ g(\mu,b) =   \gsh(\mu) + |b| ~ \textrm{sw}\left( \gsh(\mu) / |b|- \frac12 \right) $ 
with $g_{\rm sh }(\mu) = \mu / \epsilon_1$ (we assumed $\mu>0$). 
It yields strikingly different predictions compared with Eq.~\eqref{eq:g_ideal}. In particular, the conductance 
vanishes in the quantum regime, which is reached for fields 
$|b| \geq 2 \gsh  $, 
as opposed to the positive linear magneto-conductance at high fields.
Furthermore, a Lorentz broadening of the Landau levels  leads to an exponentially suppressed magneto-conductance at low field, in contrast with Eq.~\eqref{eq:largebroad} for the Weyl Hamiltonian.
%
The sharp difference between the ballistic magneto-conductance of a Weyl junction with that of a standard material with 
a non-relativistic dispersion relation is highlighted in Fig.~\ref{fig:Theory}, which illustrates the presence vs 
absence of the positive magneto-conductance in the quantum regime. 

\section{Numerical Study}
\label{sec:3}

\begin{figure}[tb]
\centerline{
    \includegraphics[width=8cm]{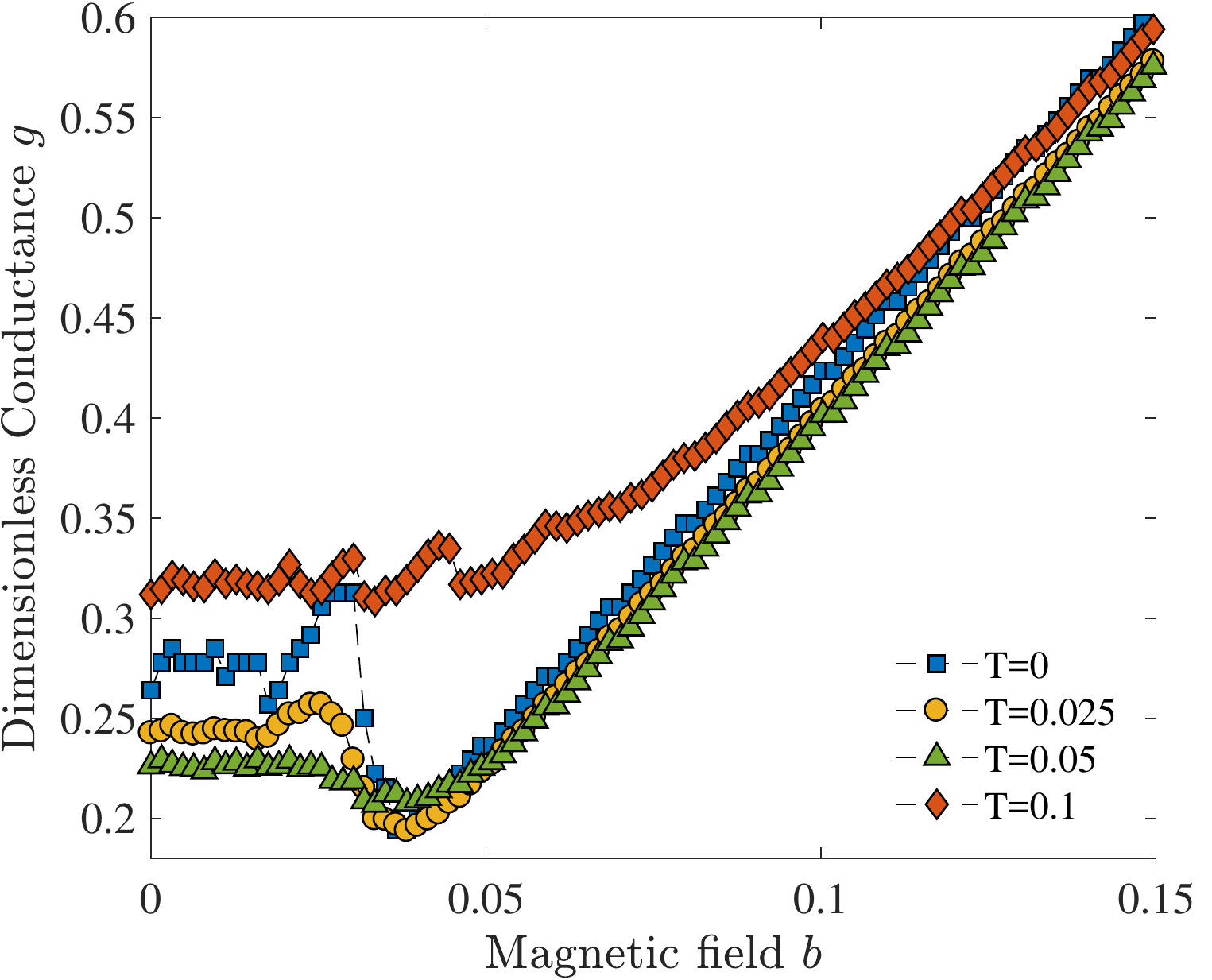}
}
	\caption{ 
	Thermal broadening of the magneto-conductance   
	The parameters are identical to those of Fig~\ref{fig_numericalConctance} with $\mu=0.6$ and 
	a transverse size $W/a=24, L/a=4$.
	\label{fig:thermalbroadening} 
	}
\end{figure}

 We now complement the previous arguments based on the simplified low-energy Bloch Hamiltonian \eqref{eq:model} by a numerical study of transport of two lattice Hamiltonians displaying respectively four pairs and one pair of Weyl cones.
First, we compute the conductance of a ballistic junction of size $W\times W\times L$  using the Kwant numerical software \cite{Groth:2014}
 applied to a tight-binding two-band model on a cubic lattice with nearest-neighbor couplings, 
such that the dispersion relation,
$E_\pm^2(\mathbf{k})=  
 t^2 ( \sin^2 k_xa  + \sin^2 k_ya ) + \left[  t_z(1-\cos k_za)-\Delta \right]^2, 
 $
 possesses four pairs of Weyl cones with opposite chiralities at quasi-wavectors $\mathbf{k}=(0/\pi , 0/\pi, \pm K_z)$ with $K_z=(2/a)\arcsin \sqrt{\Delta/(2t_z)}$, assuming $0<\Delta/(2t_z)<1$
(see Sec.~III in SM \cite{SupplMat} for details).
Throughout our study, we use  
 $a=1$ for the lattice spacing, 
$t=1$ and $t_z=1$ for the hopping matrix elements in $(xy)$-plane and along $z$-direction, respectively, and various 
values of the energy threshold $\Delta$. 
We study the transport along the $z$-direction 
as a function of the magnetic field ${\bf B}\parallel z$ aligned with the junction. 
Results for $\Delta=1$ are shown in Fig.~\ref{fig_numericalConctance} in which
 the conductance of the junction is plotted as a function the dimensionless magnetic flux $b$ 
 threading a lattice unit cell. 
For all chemical potential of the conductor below the energy threshold between Weyl valleys, $|\mu|<\Delta$, 
we find a quantitative agreement with the previous analysis. Namely, 
 (i) at low magnetic field, the conductance $g(b)$ oscillates with $b$, with an amplitude of oscillations that increases linearly with $b$,  
 (ii) at high magnetic field, the conductance $g(b)$  reaches a linear regime independent on energies, 
 $g(b) \simeq 4|b|$, where the factor $4$ 
accounts for the presence of $4$ pairs of Weyl points in our model. 
We have checked that both that the conductance is independant of the length $L$ of the junction, 
and that 
the above behavior is a contribution of bulk states, while surface states provide a subdominant 
 contribution $\propto W$ to the conductance (see SM \cite{SupplMat}, Sec.~IV).
%
Note that the numerical Landauer technique that we use amounts to introducing semi-infinite systems on both sides 
of the conducting 
part, playing the role of leads with perfect contacts,  
thereby allowing to reach a ballistic regime. In doing so we effectively consider a long ballistic junction, 
 corresponding to an artificially large dwell time for the electrons, which  
hampers the study of the algebraic low $b$ regime which is relevant experimentally. 
The thermal broadening of the magneto-conductance is illustrated in Fig.~\ref{fig:thermalbroadening}, 
in agreement with Eq.~\eqref{eq:euler-highT} for $\mu \gg kT$.  
\begin{figure}
	\includegraphics[width=8.5cm]{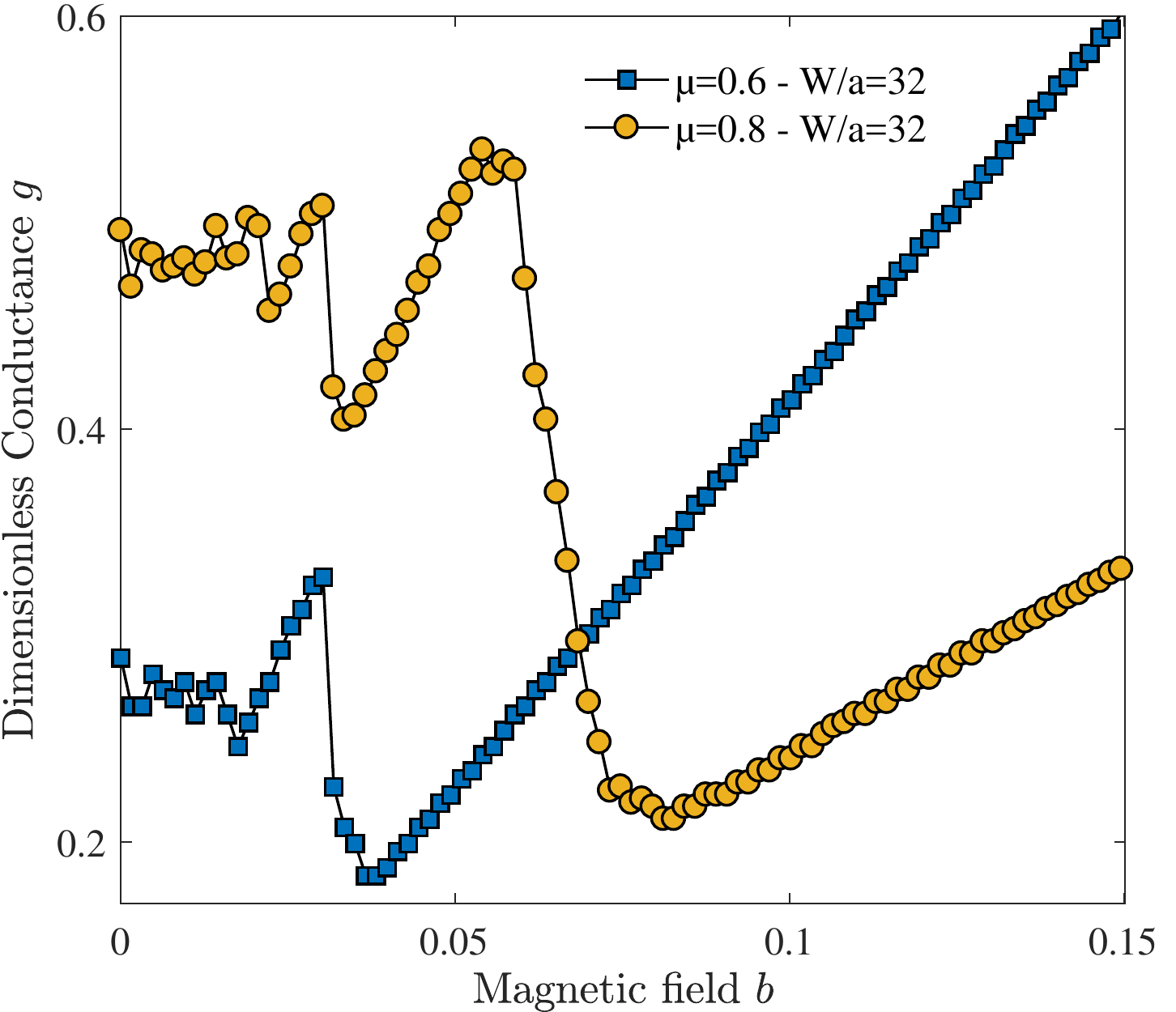}
 \caption{\label{figthib-landaulevels}
	Magneto-conductance of a model with 4 pairs of Weyl points and energy threshold $\Delta=0.7$ between them. 
 	Half of the pairs of Weyl cones have a ``chirality vector'' $\mathbf{K}$ between two Weyl cones 
 	(see Fig.~\ref{fig:DessinJunction}) 
 	that is parallel to the magnetic field, while it is antiparallel for the other half. Hence for chemical potentials above the saddle point between Weyl valleys, the 
 	linear magneto-conductance at high fields is lost for half of the pairs: the slope of the resulting 
 	magneto-conductance is reduced by a factor two.
 }
\end{figure}
\begin{figure*}[htb]
\centerline{
	\includegraphics[width=17cm]{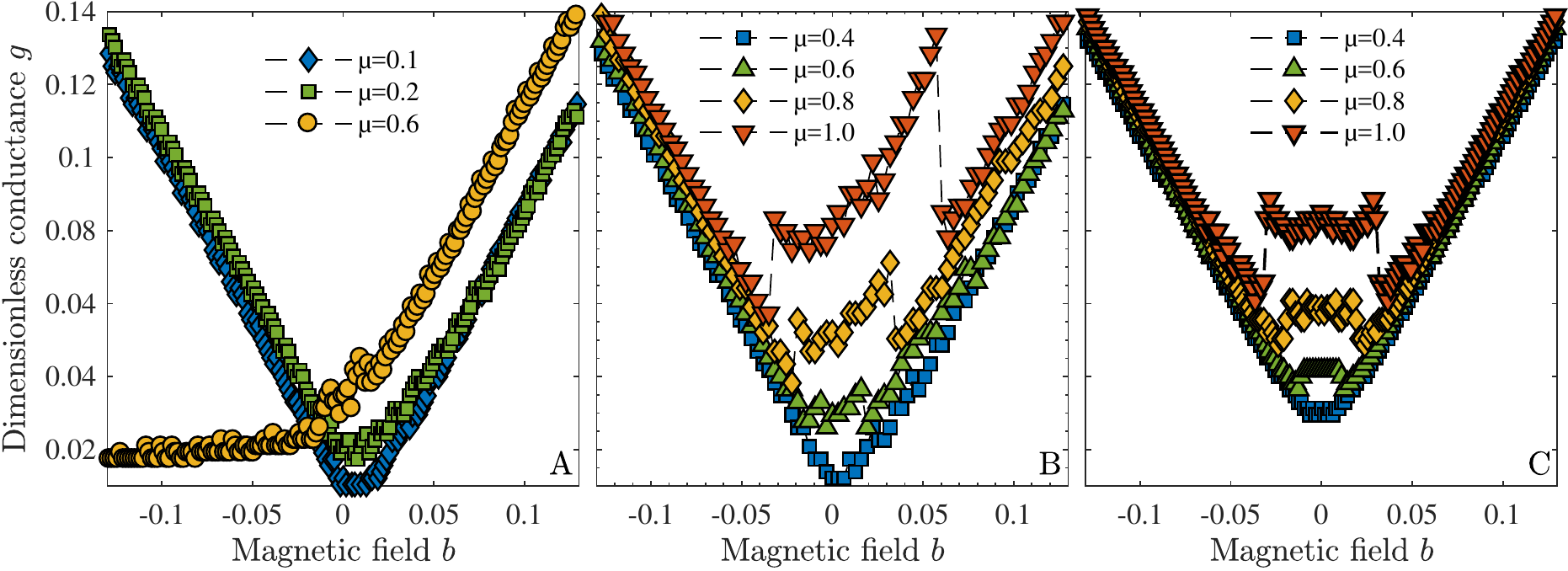}
}
	\caption{\label{fig_numericalConctance2Cones}  
	Magneto-conductance of a model with a single pair of Weyl cones separated by an energy threshold 
	$\Delta=0.3$ (A) and $\Delta=1.5$ (B,C), for a conductor size 
	$W/a=24$. 
	The linear regime at large field regime becomes asymetric for $\mu>\Delta$ as shown in Fig.~A: 
	the regime $g(b) \simeq |b|$ is lost for $b<0$. 
	Similarly, the asymmetry of the small field regime for any chemical potential $\mu$ is clearly visible for 
	$\Delta=1.5$, Fig.~B. In particular the minimum of $g(b)$ is shifted towards a positive field, 
	reflecting the breaking of time-reversal symmetry for a model with a single pair of Weyl cones located at points 
	$\pm \mathbf{K}/2$. 
	For a magnetic field transverse to this direction $\mathbf{K}$, a symetric conductance $g(b)=g(-b)$  
	is recovered as shown in Fig.~C.  
 }
\end{figure*}
\begin{figure}[htbp]
\begin{center}
	\includegraphics[width=8.5cm]{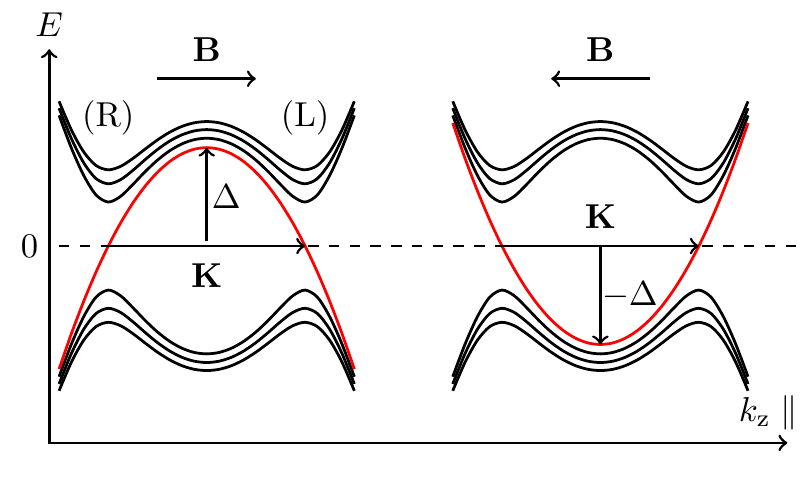}
  	\caption{\label{fig:LandauFull} 
  	Schematic view of the Landau level dispersion for a pair of left- and right-handed Weyl cones 
  	(resp. L and R) separated by an         energy threshold $\Delta$. 
   	The $n=0$ Landau band depends on the $\mathbf{B}\parallel \pm \mathbf{K}$. For 
   	$\mathbf{B}\parallel +\mathbf{K}$, this $n=0$ Landau Level exists for energies $\mu < \Delta$ 
   	while for $\mathbf{B}\parallel -\mathbf{K}$, this $n=0$ Landau Level exists for energies $\mu >- \Delta$. 
   	For a given pair or Weyl points this defines the range of chemical potential, which depends on the orientation of magnetic field, for which a ballistic linear 
   	magneto-conductance is observed at magnetic high field.  
  	}
\end{center}
\end{figure}

We now discuss the dependence of the ballistic magneto-conductance on the chemical potential, and show that 
the number of pairs of Weyl valleys contributing to this regime depends on this chemical potential. 
In particular, as 
shown in Fig.~\ref{figthib-landaulevels}, we find that the 
linear regime at large fields survives even above the threshold energy separating the Weyl valleys 
$\mu > \Delta$ (here taken as $\Delta=0.7$), though its slope is reduced by a factor two at large $\mu$.  
To understand this reduction of the slope, we consider the simpler situation of a single pair of Weyl cones 
separated by an energy threshold $\Delta$. It can be realized with a tight-binding two-band model on a cubic 
lattice introduced in Ref.~\onlinecite{Delplace:2012},
where the Weyl points only occur at $\mathbf{k}=\pm{\mathbf K}/2 = (0,0,\pm K_z)$ (see SM \cite{SupplMat}, Sec.~IV).
The corresponding numerical results for the magneto-conductance are presented in Fig.~\ref{fig_numericalConctance2Cones}. We observe that 
above the threshold energy $\Delta$ separating the Weyl valleys, the linear behavior at large field is lost for one 
direction of the magnetic field oriented along the current with 
${\mathbf B} \parallel {\mathbf I} \parallel {\mathbf K}$ 
(Figs.~\ref{fig_numericalConctance2Cones}A and \ref{fig_numericalConctance2Cones}B). 
Moreover, the whole behavior at small magnetic field is now asymmetric in $b$,
 irrespective of $\mu$. 
Note that such a simple Weyl semi-metal with only two cones necessarily breaks time-reversal symmetry (TRS).  
 This manifests itself in the breaking of Onsager relation 
$G({\bf B}) \neq G(-{\bf B}) $, and henceforth an asymmetry of the curves $G({\bf B})$ (or $g(b)$). 
 The amplitude of this TRS breaking and this asymmetry originates from 
the separation $\mathbf{K}$ of the two cones in the Brillouin zone, or more precisely on its projection 
onto the direction of magnetic field. 
Indeed,  when the vector $\mathbf{K}$ is aligned perpendicular to both the magnetic field and the junction, 
no sign of this TRS breaking is observed on magneto-transport and 
a magneto-conductance symmetric in $b$ is recovered as shown in Fig.~\ref{fig_numericalConctance2Cones}C.

The change of linear magneto-conductance in the quantum regime 
can be understood by considering the simple Bloch Hamiltonian that generalizes Eq.~\eqref{eq:model} and describes 
two Weyl valleys separated by an energy saddle point,
\begin{equation}
\label{eq:modelDelta}
H = 
c p_x \sigma_x +  cp_y \sigma_y + 
\left( \frac{p_z^2}{2m} - \Delta \right) \sigma_z .
\end{equation}
For $\Delta >0$, two Weyl points at $\pm \mathbf{K}/2 =(0,0, \pm \sqrt{2m \Delta}/\hbar)$, and having opposite
 chirality, are separated by an energy barrier $\Delta$. 
The chemical potential  dependance in the large field regime can now be inferred from the behavior of the Landau 
levels for the model
\eqref{eq:modelDelta} (see SM \cite{SupplMat}, Sec.~I). 
Depending on the sign of $b$, i.e., on the orientation of the field with respect to the ``chirality vector'' $\mathbf{K}$ 
pointing from the left-handed Weyl cone with $\chi=-1$ to the right-handed Weyl cone with $\chi=+1$, 
the dispersion of the $n=0$ Landau level changes dramatically. Indeed, it either exists for $\mu$ smaller than 
$\Delta$, 
or for $\mu$ larger than $-\Delta$ as illustrated in Fig.~\ref{fig:LandauFull}. 
We expect the larger energy barrier between the two Weyl valleys, neglected in the model \eqref{eq:modelDelta}, 
 to cut-off the energy range of this $n=0$ Landau band on the other side.
Beyond this saddle point energy, the linear
magneto-conductance for the pair of Weyl cones is lost. Depending on the sign of $b$, this happens either above the
positive or below the negative energy saddle point $\pm \Delta$. 

We are now ready to understand the reduction of the slope illustrated in Fig.~\ref{figthib-landaulevels}  for the model with four pairs of Weyl cones. Indeed, due to the symmetrical distribution of the pairs of Weyl cones in 
momentum space in that model, above the saddle point energy,
the linear magneto-conductance contribution vanishes for the half of Weyl pairs whose chirality vector is parallel 
with the magnetic field, while it persists for the other half having an 
anti-parallel chirality vector with respect to $\mathbf{b}$. 
Furthermore, the absence of an asymmetry of the magneto-conductance in that model can also be traced to the 
symmetrical distribution of Weyl cones in momentum space.

%

\begin{figure}
	\includegraphics[width=8.5cm]{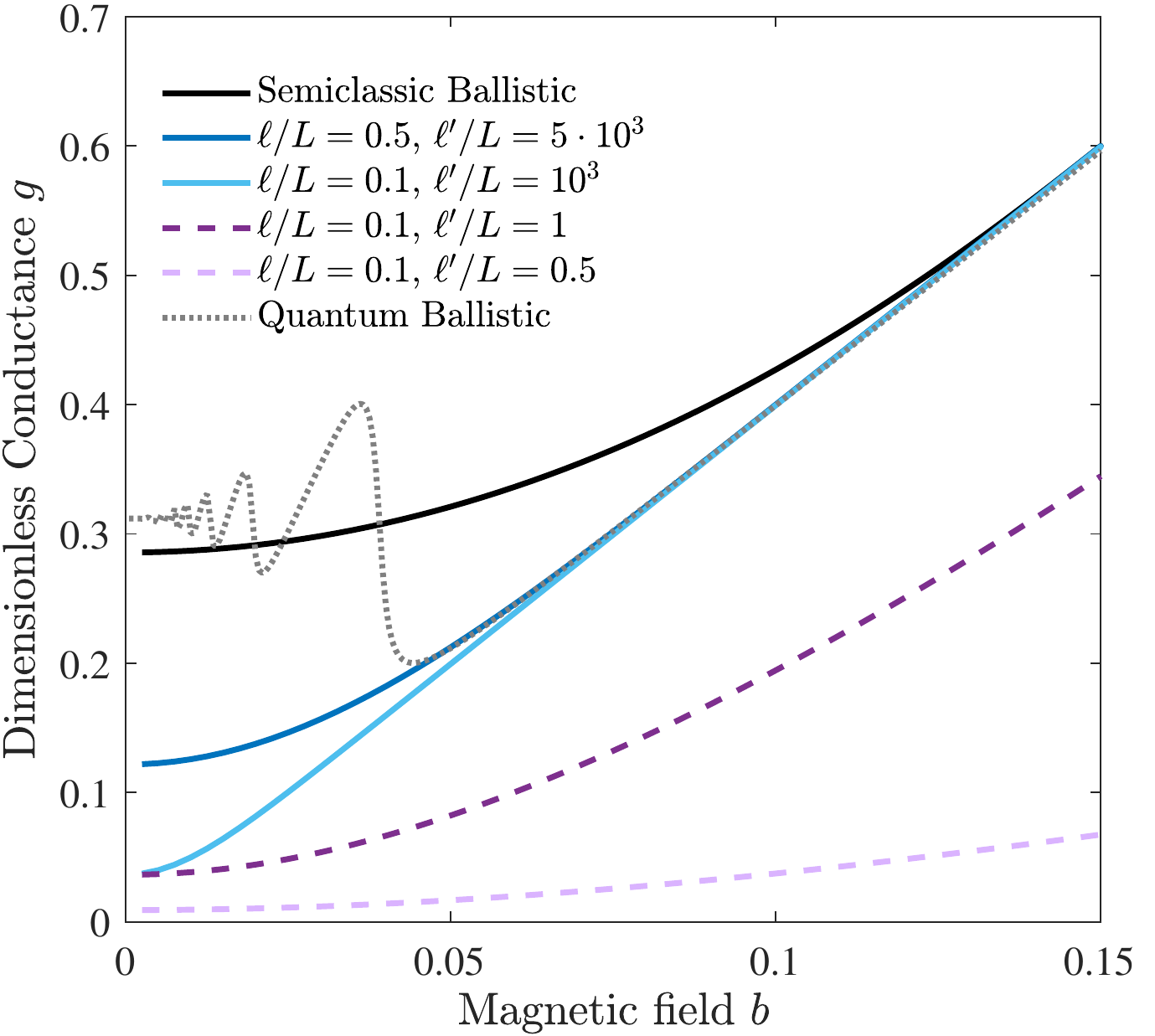}
 \caption{\label{fig:SemiClassic}
	Semi-classical magneto-conductance  $g(b)$ of a Weyl junction 
 for different inter and intra cone scattering lengths $\ell,\ell'$ and $\mu = 0.7$. 
The black plain line corresponds to the ballistic junction with $L \leq \ell, \ell'$, and displays a $b^2$-correction for weak 
magnetic fields while a linear quantum magneto-conductance $g(b)=|b|$ is recovered at large fields. 
When only the  intra-cone scattering length is smaller than the junction length,  $\ell \leq L \leq \ell'$, a diffusive regime for 
cold electrons is found at small fields while the  linear quantum regime is unaffected and reached at smaller fields as represented by the plain blue curves. 
Finally when $\ell \leq \ell'  \leq L$ the linear regime at high fields only survives at very high magnetic fields. 
 }
\end{figure}

\section{Semi-classical description of ballistic transport}
\label{sec:4}

Let us now describe within a semi-classical picture the above ballistic magneto-transport. 
The advantage of such an approach is that it allows describing the smooth 
cross-over  from the ballistic to the diffusive regime, as well as compare the results in both regimes. 
In this approach, the evolution of a semi-classical wave-packet of Weyl eigenstates is described by the 
semi-classical equations of motion (for a vanishing electric field) \cite{Xiao:2010}: 
\begin{equation}
\partial_{t}\mathbf{x} = \frac{1}{\hbar} 
    \partial_{\mathbf{k}} \varepsilon - \partial_{t}\mathbf{k} \times \mathbf{\Omega} 
\ ;\
\hbar \partial_{t}\mathbf{k} =  - e \partial_{t} \mathbf{x} \times \mathbf{B} . 
\label{eq:semiclassicalEOM}
\end{equation}
Here, $\mathbf{x}$ and $\mathbf{k}$ are the position and momentum of the wavepacket, $\varepsilon_{\mathbf{ k}}$ and $ \Omega^{\chi}_{\mathbf{ k}}$ are the energy and  Berry curvature for the conduction band, which, around a Weyl point 
described by a Hamiltonian 
$ H^\chi_{\mathbf{ k}} = \hbar v_F \chi ~\boldsymbol{\sigma} \cdot {\mathbf{ k}}$,  read $\epsilon_{\mathbf{ k}}=\hbar v_F k$ and 
$
 \Omega^{\chi}_{\mathbf{ k}} = \chi{\mathbf{ k}} /(2 |\mathbf{k}|^3). 
$
 Solving for the semi-classical equations of motion \eqref{eq:semiclassicalEOM}, 
we obtain the anomalous  velocity of the electrons, corrected 
by the Berry curvature:  
$\mathbf{w}_{\chi}  = \partial_{\mathbf{ k}} \varepsilon + (\partial_{\mathbf{ k}} \varepsilon \cdot \boldsymbol{\Omega}) 
{\bf B} $. 
The charge current density along the $z$-direction follows: 
\begin{equation}
 j_z =  - e \sum_{\chi,\hat{ k}} (\mathbf{w}_{\chi}  \cdot \hat z )~  f_\chi (\mathbf{ x} , \epsilon, \hat{ k}),
\label{eq:current}
\end{equation}
where the sum runs over constant energy contours and $\hat{k} = \mathbf{k}/|\mathbf{k}|$. 
The distribution function satisfies the stationary Boltzmann equation (dropping the energy dependance of $f$) 
\begin{multline}
	(\mathbf{w}_{\chi}.\hat{z}) ~\partial_z f_{\chi}(z,\hat{k})=
 \frac{1}{\tau}\sum_{\hat{k}'}\left[f_{\chi}(z,\hat{k}')-f_{\chi}(z,\hat{k})\right]
 \\
 +\frac{1}{\tau'}\sum_{\hat{k}'}\left[f_{-\chi}(z,\hat{k}')-f_{\chi}(z,\hat{k})\right] . 
\label{eq:kinetic0}
\end{multline}
Here $\tau$ and $\tau'$ are the intra-cone and inter-cone elastic scattering times, respectively, 
which are assumed to be 
larger than the dwell time in the ballistic regime.  
 We solve Eq.~\eqref{eq:kinetic0} with an ansatz for $f$ that satisfies boundary conditions at the contacts 
 \cite{Jong:1994},
\begin{equation}
 f = f_0 (\varepsilon_{\mathbf{k}} - \mu_1) \Theta \left[ \mathbf{w}_{\chi}  \cdot \hat z \right] 
 + f_0 (\varepsilon_{\mathbf{k}} - \mu_2) \Theta \left[ - \mathbf{w}_{\chi}  \cdot \hat z \right]. 
\end{equation}
Solving these equations in the ballistic regime, $\tau, \tau' \to \infty$, we find 
\begin{equation}
g_{cl} = \left\{
	\begin{array}{ll}
	\gsh \left[ 1 +   b^2 / (2\gsh)^2 \right] & \textrm{for } |b| < 2 \gsh, \\
	|b|  & \textrm{for } |b| \geq 2  \gsh .
	\end{array}
	\right.
\end{equation}

Quite remarkably, incorporating the Berry curvature effect into the semi-classical description of the ballistic
 transport allows to describe the anomalous linear magneto-conductance in the quantum regime at large field, 
as well as the cross-over towards 
an anomalous regime at low field. However, the semi-classical description is unable to accurately describe the 
$|b|^{3/2}$-dependence of the magneto-conductance, see Eq.~\eqref{eq:largebroad}, 
and predicts a $b^2$-behavior, similar to the diffusive regime \cite{Son:2013}.  
 Such a diffusive regime is indeed recovered in the present situation at finite scattering times $\tau,\tau'$. However it 
 corresponds to a different situation from the one considered by Son et Spivak \cite{Son:2013}, as inelastic 
scattering occurs in the leads in our case, and not in the junction, thus yielding a quantitatively different result.
Solving this problem, we find an 
expression for the conductance (see Sec.~VI in SM \cite{SupplMat}), which identifies with that previously 
derived {\it via} a topological non-linear sigma field theory~\cite{Altland:2016}. 
As expected, when intra-cone disorder is increased a diffusive regime is reached  at small magnetic field 
when $\ell = c\tau \leq L$, 
with a conductance now scaling as $G \propto W^2 /L$. The correction at small field in this regime remains 
quadratic in $b$. At high magnetic field, the previous ballistic quantum magneto-conductance 
regime $g(b) = |b|$ is affected neither by  intra-cone, nor by inter-cone disorder. 
Moreover as long as $\ell \ell' \gg L^2$, including the situation when only the intra-cone disorder is 
relevant
 $\ell \leq L \leq  \ell' = c \tau'$,  
this ballistic linear regime is reached at smaller magnetic fields, $|b| \geq   \gsh  \ell /L $ \cite{SupplMat}. 
On the other hand, when $L^2 \gg \ell \ell' $, or $L\geq \ell' \geq \ell$,  
 inter-cone disorder pushes this ballistic regime to high magnetic field (possibly 
outside of the experimental regime) for 
$|b| \geq   \gsh  L/\ell' $. 
 The magneto-conductance corresponding to these different regimes is obtained by numerically solving 
 the semi-classical 
 diffusive equation (see \cite{SupplMat}, Sec.~VI), and the results are represented in 
 Fig.~\ref{fig:SemiClassic}.

\section{Discussion}
\label{sec:5}

In this paper, we have shown that two regimes have to be distinguished when discussing the ballistic conductance of a Weyl junction in parallel magnetic field  : 
(i) a weak magnetic field regime, in which the magneto-conductance behaves algebraically with the magnetic field, whose details are non-universal. Furthermore at higher fields the magneto-conductance displays 
quantum oscillations  when the broadening of Landau levels is weak (such as in long ballistic junctions) 
and at low temperature,
(ii) a linear regime at high fields with a universal slope. 

Let us now discuss the relation between these different regimes and the underlying chiral anomaly. 
In the ballistic regime addressed in this work, there exists an \emph{equilibrium chiral current} density 
$j_5 = j_R - j_L \propto |b|$ for non vanishing magnetic fields: the presence of the $n=0$ Landau band implies
 that, irrespective of the chemical potential $\mu$, there is an excess of Weyl electrons of, {\it e.g.}, left-chirality moving 
 to the right of the junction, and electrons of right-chirality moving to the left. 
  This equilibrium chiral current is obviously uncorrelated with a  charge density current $j = j_R + j_L$, as the later   
  vanishes in equilibrium. 
 On the other hand, 
in the presence of a chemical potential bias, $\mu_1 - \mu_2 = \delta \mu$, the excess of chiral current, or 
\emph{non-equilibrium chiral current}, can manifest itself in a charge current, as first discussed in Ref.~\cite{Nielsen:1983}. 
This non-equilibrium chiral current is still entirely due to the $n=0$ Landau band and 
reads $e^2/h (W/a)^2 |b| (\delta \mu /e)$. 
At weak magnetic field only part of the conductance \eqref{eq:g_b} is related to the 
$n=0$ Landau band contribution, and thus can be related to the chiral anomaly. 
 Hence the anomalous positive ballistic magneto-conductance in the low field regime is not a unique signature of the chiral 
 anomaly. 
 This is in contrast with the situation of "hot electrons", close to equilibrium, considered in Ref.~\cite{Son:2013}. There,  
 the chiral current driven by the $n=0$ Landau level leads to a chiral chemical potential $\mu_R - \mu_L $ 
 between left- and right-handed Weyl valleys, and a magneto-conductance via the chiral magnetic effect. 
 Hence attributing the anomalous magneto-conductance at low fields to the chiral anomaly requires first to identify 
 the relevant regime of transport, and the observation of a positive algebraic behavior is not sufficient. 

In contrast to the low field regime, 
the quantum regime of a linear magneto-conductance $G = W^2 (e^3/h^2)|B| $ 
at high field is a unique contribution of the $n=0$ 
Landau level. Thus it can be unambiguously associated with the chiral anomaly. 
This regime is reached for magnetic fields satisfying 
$B \lambda^2_F \geq h/e$ where $\lambda_F$ is the Fermi wavelength. 
Thus it can be reached experimentally for chemical potentials sufficiently close to the band crossing. 
Moreover this regime is robust, 
and persists even in the presence of disorder: only its domain of existence is affected by elastic scattering. 
Furthermore, when energy relaxation occurs within the conductor the above ballistic conductance at high fields
is replaced by $G=W^2 (\ell'/L)  (e^3/h^2)|B| $, 
also linear in 
magnetic field \cite{Nielsen:1983,Aji:2012,Gorbar:2014}. In this regime the slope of the 
linear regime now depends explicitly on the amplitude of inter-cone scattering. Hence the study of this linear 
regime should provide an unambiguous determination of the regime of transport. In the exact same 
regime of transport a linear magneto-resistance in transverse magnetic field was also predicted by Abrikosov in Refs.~\cite{Abrikosov:1998,Abrikosov:2000} (see also \cite{Lu:2015}).  
We believe that this quantum regime of transport in Weyl materials is of high experimental interest. 
Quite remarkably a linear-in-field longitudinal magnetic conductance has already 
been observed  in narrow wires of NbP, a Weyl semi-metal \cite{Niemann:2017}, hence validating that this regime in within experimental reach, although it hasn't been thouroughly studied.

\section{Conclusion}
\label{sec:6}

 In this paper, we have studied the conductance of a junction of Weyl material in the presence of a parallel magnetic field, and in the ballistic regime. We have shown that the low-field magneto-conductance displays low-temperature quantum oscillations, whose broadening 
 results in an algebraic behavior at vanishing field. 
 At high fields, the magneto-conductance becomes linear in the field. Besides its experimental relevance, this ballistic regime allows discussing  
in details the relation between this conductance and the chiral anomaly of Weyl fermions. This allows to unambiguously identify  the large field regime as a signature of the chiral anomaly.

\section*{Acknowledgments}
We would like to thank Gwendal F{\` e}ve, Jean-No{\"e}l Fuchs, Mark Goerbig, Fr{\'e}d{\'e}ric Pi{\'e}chon, 
Bernard Pla{\c c}ais  for stimulating discussions and comments on the manuscrit. 
D.C. and M. H. thank the International Institute of Physics (Natal, Brazil) where this project was initiated 
and  
D.C. thanks the  LPA at ENS-Paris for hospitality during the last stage of this work. 

\begin{thebibliography}{32}%
\makeatletter
\providecommand \@ifxundefined [1]{%
 \@ifx{#1\undefined}
}%
\providecommand \@ifnum [1]{%
 \ifnum #1\expandafter \@firstoftwo
 \else \expandafter \@secondoftwo
 \fi
}%
\providecommand \@ifx [1]{%
 \ifx #1\expandafter \@firstoftwo
 \else \expandafter \@secondoftwo
 \fi
}%
\providecommand \natexlab [1]{#1}%
\providecommand \enquote  [1]{``#1''}%
\providecommand \bibnamefont  [1]{#1}%
\providecommand \bibfnamefont [1]{#1}%
\providecommand \citenamefont [1]{#1}%
\providecommand \href@noop [0]{\@secondoftwo}%
\providecommand \href [0]{\begingroup \@sanitize@url \@href}%
\providecommand \@href[1]{\@@startlink{#1}\@@href}%
\providecommand \@@href[1]{\endgroup#1\@@endlink}%
\providecommand \@sanitize@url [0]{\catcode `\\12\catcode `\$12\catcode
  `\&12\catcode `\#12\catcode `\^12\catcode `\_12\catcode `\%12\relax}%
\providecommand \@@startlink[1]{}%
\providecommand \@@endlink[0]{}%
\providecommand \url  [0]{\begingroup\@sanitize@url \@url }%
\providecommand \@url [1]{\endgroup\@href {#1}{\urlprefix }}%
\providecommand \urlprefix  [0]{URL }%
\providecommand \Eprint [0]{\href }%
\providecommand \doibase [0]{http://dx.doi.org/}%
\providecommand \selectlanguage [0]{\@gobble}%
\providecommand \bibinfo  [0]{\@secondoftwo}%
\providecommand \bibfield  [0]{\@secondoftwo}%
\providecommand \translation [1]{[#1]}%
\providecommand \BibitemOpen [0]{}%
\providecommand \bibitemStop [0]{}%
\providecommand \bibitemNoStop [0]{.\EOS\space}%
\providecommand \EOS [0]{\spacefactor3000\relax}%
\providecommand \BibitemShut  [1]{\csname bibitem#1\endcsname}%
\let\auto@bib@innerbib\@empty
\bibitem [{\citenamefont {Adler}(1969)}]{Adler:1969}%
  \BibitemOpen
  \bibfield  {author} {\bibinfo {author} {\bibfnamefont {S.}~\bibnamefont
  {Adler}},\ }\href@noop {} {\bibfield  {journal} {\bibinfo  {journal} {Phys.
  Rev.}\ }\textbf {\bibinfo {volume} {177}},\ \bibinfo {pages} {2426} (\bibinfo
  {year} {1969})}\BibitemShut {NoStop}%
\bibitem [{\citenamefont {Bell}\ and\ \citenamefont
  {Jackiw}(1969)}]{Bell:1969}%
  \BibitemOpen
  \bibfield  {author} {\bibinfo {author} {\bibfnamefont {J.~S.}\ \bibnamefont
  {Bell}}\ and\ \bibinfo {author} {\bibfnamefont {R.}~\bibnamefont {Jackiw}},\
  }\href@noop {} {\bibfield  {journal} {\bibinfo  {journal} {Nuovo Cimento A}\
  }\textbf {\bibinfo {volume} {60}},\ \bibinfo {pages} {47} (\bibinfo {year}
  {1969})}\BibitemShut {NoStop}%
\bibitem [{\citenamefont {Kharzeev}(2014)}]{Kharzeev:2014}%
  \BibitemOpen
  \bibfield  {author} {\bibinfo {author} {\bibfnamefont {D.}~\bibnamefont
  {Kharzeev}},\ }\href@noop {} {\bibfield  {journal} {\bibinfo  {journal}
  {Progr. Part. Nucl. Phys.}\ }\textbf {\bibinfo {volume} {75}},\ \bibinfo
  {pages} {133} (\bibinfo {year} {2014})}\BibitemShut {NoStop}%
\bibitem [{\citenamefont {Herring}(1937)}]{Herring:1937}%
  \BibitemOpen
  \bibfield  {author} {\bibinfo {author} {\bibfnamefont {C.}~\bibnamefont
  {Herring}},\ }\href@noop {} {\bibfield  {journal} {\bibinfo  {journal} {Phys.
  Rev.}\ }\textbf {\bibinfo {volume} {52}},\ \bibinfo {pages} {365} (\bibinfo
  {year} {1937})}\BibitemShut {NoStop}%
\bibitem [{\citenamefont {Volovik}(2003)}]{Volovik:2003}%
  \BibitemOpen
  \bibfield  {author} {\bibinfo {author} {\bibfnamefont {G.~E.}\ \bibnamefont
  {Volovik}},\ }\href@noop {} {\emph {\bibinfo {title} {The Universe in a
  Helium Droplet}}}\ (\bibinfo  {publisher} {Clarendon Press},\ \bibinfo {year}
  {2003})\BibitemShut {NoStop}%
\bibitem [{\citenamefont {Wan}\ \emph {et~al.}(2011)\citenamefont {Wan},
  \citenamefont {Turner}, \citenamefont {Vishwanath},\ and\ \citenamefont
  {Savrasov}}]{Wan:2011}%
  \BibitemOpen
  \bibfield  {author} {\bibinfo {author} {\bibfnamefont {X.}~\bibnamefont
  {Wan}}, \bibinfo {author} {\bibfnamefont {A.~M.}\ \bibnamefont {Turner}},
  \bibinfo {author} {\bibfnamefont {A.}~\bibnamefont {Vishwanath}}, \ and\
  \bibinfo {author} {\bibfnamefont {S.~Y.}\ \bibnamefont {Savrasov}},\
  }\href@noop {} {\bibfield  {journal} {\bibinfo  {journal} {Phys. Rev. B}\
  }\textbf {\bibinfo {volume} {83}},\ \bibinfo {pages} {205101} (\bibinfo
  {year} {2011})}\BibitemShut {NoStop}%
\bibitem [{\citenamefont {Bansil}\ \emph {et~al.}(2016)\citenamefont {Bansil},
  \citenamefont {Lin},\ and\ \citenamefont {Das}}]{Bansil:2016}%
  \BibitemOpen
  \bibfield  {author} {\bibinfo {author} {\bibfnamefont {A.}~\bibnamefont
  {Bansil}}, \bibinfo {author} {\bibfnamefont {H.}~\bibnamefont {Lin}}, \ and\
  \bibinfo {author} {\bibfnamefont {T.}~\bibnamefont {Das}},\ }\href@noop {}
  {\bibfield  {journal} {\bibinfo  {journal} {Rev. Mod. Phys.}\ }\textbf
  {\bibinfo {volume} {88}},\ \bibinfo {pages} {021004} (\bibinfo {year}
  {2016})}\BibitemShut {NoStop}%
\bibitem [{\citenamefont {Nielsen}\ and\ \citenamefont
  {Ninomiya}(1983)}]{Nielsen:1983}%
  \BibitemOpen
  \bibfield  {author} {\bibinfo {author} {\bibfnamefont {H.}~\bibnamefont
  {Nielsen}}\ and\ \bibinfo {author} {\bibfnamefont {M.}~\bibnamefont
  {Ninomiya}},\ }\href {\doibase
  http://dx.doi.org/10.1016/0370-2693(83)91529-0} {\bibfield  {journal}
  {\bibinfo  {journal} {Physics Letters B}\ }\textbf {\bibinfo {volume}
  {130}},\ \bibinfo {pages} {389 } (\bibinfo {year} {1983})}\BibitemShut
  {NoStop}%
\bibitem [{\citenamefont {Aji}(2012)}]{Aji:2012}%
  \BibitemOpen
  \bibfield  {author} {\bibinfo {author} {\bibfnamefont {V.}~\bibnamefont
  {Aji}},\ }\href@noop {} {\bibfield  {journal} {\bibinfo  {journal} {Phys.
  Rev. B}\ }\textbf {\bibinfo {volume} {85}},\ \bibinfo {pages} {241101}
  (\bibinfo {year} {2012})},\ \bibinfo {note} {arXiv:1108.4426}\BibitemShut
  {NoStop}%
\bibitem [{\citenamefont {Son}\ and\ \citenamefont {Spivak}(2013)}]{Son:2013}%
  \BibitemOpen
  \bibfield  {author} {\bibinfo {author} {\bibfnamefont {D.~T.}\ \bibnamefont
  {Son}}\ and\ \bibinfo {author} {\bibfnamefont {B.~Z.}\ \bibnamefont
  {Spivak}},\ }\href@noop {} {\bibfield  {journal} {\bibinfo  {journal} {Phys.
  Rev. B}\ }\textbf {\bibinfo {volume} {88}},\ \bibinfo {pages} {104412}
  (\bibinfo {year} {2013})}\BibitemShut {NoStop}%
\bibitem [{\citenamefont {Burkov}(2014)}]{Burkov:2014}%
  \BibitemOpen
  \bibfield  {author} {\bibinfo {author} {\bibfnamefont {A.}~\bibnamefont
  {Burkov}},\ }\href@noop {} {\bibfield  {journal} {\bibinfo  {journal} {Phys.
  Rev. Lett.}\ }\textbf {\bibinfo {volume} {113}},\ \bibinfo {pages} {247203}
  (\bibinfo {year} {2014})}\BibitemShut {NoStop}%
\bibitem [{\citenamefont {Burkov}(2015)}]{Burkov:2015}%
  \BibitemOpen
  \bibfield  {author} {\bibinfo {author} {\bibfnamefont {A.}~\bibnamefont
  {Burkov}},\ }\href@noop {} {\bibfield  {journal} {\bibinfo  {journal}
  {Journal of Physics: Condensed Matter}\ }\textbf {\bibinfo {volume} {27}},\
  \bibinfo {pages} {3201} (\bibinfo {year} {2015})}\BibitemShut {NoStop}%
\bibitem [{\citenamefont {Chen}\ \emph {et~al.}(2013)\citenamefont {Chen},
  \citenamefont {Wu},\ and\ \citenamefont {Burkov}}]{Chen:2013}%
  \BibitemOpen
  \bibfield  {author} {\bibinfo {author} {\bibfnamefont {Y.}~\bibnamefont
  {Chen}}, \bibinfo {author} {\bibfnamefont {S.}~\bibnamefont {Wu}}, \ and\
  \bibinfo {author} {\bibfnamefont {A.~A.}\ \bibnamefont {Burkov}},\ }\href
  {\doibase 10.1103/PhysRevB.88.125105} {\bibfield  {journal} {\bibinfo
  {journal} {Phys. Rev. B}\ }\textbf {\bibinfo {volume} {88}},\ \bibinfo
  {pages} {125105} (\bibinfo {year} {2013})}\BibitemShut {NoStop}%
\bibitem [{\citenamefont {Zhong}\ \emph {et~al.}(2016)\citenamefont {Zhong},
  \citenamefont {Moore},\ and\ \citenamefont {Souza}}]{Zhong:2016}%
  \BibitemOpen
  \bibfield  {author} {\bibinfo {author} {\bibfnamefont {S.}~\bibnamefont
  {Zhong}}, \bibinfo {author} {\bibfnamefont {J.~E.}\ \bibnamefont {Moore}}, \
  and\ \bibinfo {author} {\bibfnamefont {I.}~\bibnamefont {Souza}},\ }\href
  {\doibase 10.1103/PhysRevLett.116.077201} {\bibfield  {journal} {\bibinfo
  {journal} {Phys. Rev. Lett.}\ }\textbf {\bibinfo {volume} {116}},\ \bibinfo
  {pages} {077201} (\bibinfo {year} {2016})}\BibitemShut {NoStop}%
\bibitem [{\citenamefont {Baireuther}\ \emph {et~al.}(2015)\citenamefont
  {Baireuther}, \citenamefont {Hutasoit}, \citenamefont {Tworzydlo},\ and\
  \citenamefont {Beenakker}}]{Baireuther:2015}%
  \BibitemOpen
  \bibfield  {author} {\bibinfo {author} {\bibfnamefont {P.}~\bibnamefont
  {Baireuther}}, \bibinfo {author} {\bibfnamefont {J.}~\bibnamefont
  {Hutasoit}}, \bibinfo {author} {\bibfnamefont {J.}~\bibnamefont {Tworzydlo}},
  \ and\ \bibinfo {author} {\bibfnamefont {C.~W.~J.}\ \bibnamefont
  {Beenakker}},\ }\href@noop {} {\bibfield  {journal} {\bibinfo  {journal} {New
  Journal of Physics}\ }\textbf {\bibinfo {volume} {18}} (\bibinfo {year}
  {2015})}\BibitemShut {NoStop}%
\bibitem [{\citenamefont {Li}\ \emph {et~al.}(2015)\citenamefont {Li},
  \citenamefont {He}, \citenamefont {Lu}, \citenamefont {Zhang}, \citenamefont
  {Liu}, \citenamefont {Ma}, \citenamefont {Fan}, \citenamefont {Shen},\ and\
  \citenamefont {Wang}}]{Li:2015}%
  \BibitemOpen
  \bibfield  {author} {\bibinfo {author} {\bibfnamefont {H.}~\bibnamefont
  {Li}}, \bibinfo {author} {\bibfnamefont {H.}~\bibnamefont {He}}, \bibinfo
  {author} {\bibfnamefont {H.-Z.}\ \bibnamefont {Lu}}, \bibinfo {author}
  {\bibfnamefont {H.}~\bibnamefont {Zhang}}, \bibinfo {author} {\bibfnamefont
  {H.}~\bibnamefont {Liu}}, \bibinfo {author} {\bibfnamefont {R.}~\bibnamefont
  {Ma}}, \bibinfo {author} {\bibfnamefont {Z.}~\bibnamefont {Fan}}, \bibinfo
  {author} {\bibfnamefont {S.-Q.}\ \bibnamefont {Shen}}, \ and\ \bibinfo
  {author} {\bibfnamefont {J.}~\bibnamefont {Wang}},\ }\href@noop {} {\bibfield
   {journal} {\bibinfo  {journal} {Nat. Comm.}\ }\textbf {\bibinfo {volume}
  {7}},\ \bibinfo {pages} {10301} (\bibinfo {year} {2015})}\BibitemShut
  {NoStop}%
\bibitem [{\citenamefont {Xiong}\ \emph {et~al.}(2015)\citenamefont {Xiong},
  \citenamefont {Kushwaha}, \citenamefont {Liang}, \citenamefont {Krizan},
  \citenamefont {Hirschberger}, \citenamefont {Wang}, \citenamefont {Cava},\
  and\ \citenamefont {Ong}}]{Xiong:2015}%
  \BibitemOpen
  \bibfield  {author} {\bibinfo {author} {\bibfnamefont {J.}~\bibnamefont
  {Xiong}}, \bibinfo {author} {\bibfnamefont {S.~K.}\ \bibnamefont {Kushwaha}},
  \bibinfo {author} {\bibfnamefont {T.}~\bibnamefont {Liang}}, \bibinfo
  {author} {\bibfnamefont {J.~W.}\ \bibnamefont {Krizan}}, \bibinfo {author}
  {\bibfnamefont {M.}~\bibnamefont {Hirschberger}}, \bibinfo {author}
  {\bibfnamefont {W.}~\bibnamefont {Wang}}, \bibinfo {author} {\bibfnamefont
  {R.~J.}\ \bibnamefont {Cava}}, \ and\ \bibinfo {author} {\bibfnamefont
  {N.~P.}\ \bibnamefont {Ong}},\ }\href@noop {} {\bibfield  {journal} {\bibinfo
   {journal} {Science}\ }\textbf {\bibinfo {volume} {350}},\ \bibinfo {pages}
  {6259} (\bibinfo {year} {2015})}\BibitemShut {NoStop}%
\bibitem [{\citenamefont {Huang}\ \emph {et~al.}(2015)\citenamefont {Huang},
  \citenamefont {Zhao}, \citenamefont {Long}, \citenamefont {Wang},
  \citenamefont {Chen}, \citenamefont {Yang}, \citenamefont {Liang},
  \citenamefont {Xue}, \citenamefont {Weng}, \citenamefont {Fang},
  \citenamefont {Dai},\ and\ \citenamefont {Chen}}]{Huang:2015}%
  \BibitemOpen
  \bibfield  {author} {\bibinfo {author} {\bibfnamefont {X.}~\bibnamefont
  {Huang}}, \bibinfo {author} {\bibfnamefont {L.}~\bibnamefont {Zhao}},
  \bibinfo {author} {\bibfnamefont {Y.}~\bibnamefont {Long}}, \bibinfo {author}
  {\bibfnamefont {P.}~\bibnamefont {Wang}}, \bibinfo {author} {\bibfnamefont
  {D.}~\bibnamefont {Chen}}, \bibinfo {author} {\bibfnamefont {Z.}~\bibnamefont
  {Yang}}, \bibinfo {author} {\bibfnamefont {H.}~\bibnamefont {Liang}},
  \bibinfo {author} {\bibfnamefont {M.}~\bibnamefont {Xue}}, \bibinfo {author}
  {\bibfnamefont {H.}~\bibnamefont {Weng}}, \bibinfo {author} {\bibfnamefont
  {Z.}~\bibnamefont {Fang}}, \bibinfo {author} {\bibfnamefont {X.}~\bibnamefont
  {Dai}}, \ and\ \bibinfo {author} {\bibfnamefont {G.}~\bibnamefont {Chen}},\
  }\href@noop {} {\bibfield  {journal} {\bibinfo  {journal} {Phys. Rev. X}\
  }\textbf {\bibinfo {volume} {5}},\ \bibinfo {pages} {031023} (\bibinfo {year}
  {2015})}\BibitemShut {NoStop}%
\bibitem [{\citenamefont {Niemann}\ \emph {et~al.}(2016)\citenamefont
  {Niemann}, \citenamefont {Gooth}, \citenamefont {Wu}, \citenamefont
  {B{\"a}{\ss}ler}, \citenamefont {Sergelius}, \citenamefont {H{\"u}hne},
  \citenamefont {Rellinghaus}, \citenamefont {Shekhar}, \citenamefont
  {S{\"u}{\ss}}, \citenamefont {Schmidt}, \citenamefont {Felser}, \citenamefont
  {Yan},\ and\ \citenamefont {Nielsch}}]{Niemann:2017}%
  \BibitemOpen
  \bibfield  {author} {\bibinfo {author} {\bibfnamefont {A.}~\bibnamefont
  {Niemann}}, \bibinfo {author} {\bibfnamefont {J.}~\bibnamefont {Gooth}},
  \bibinfo {author} {\bibfnamefont {S.-C.}\ \bibnamefont {Wu}}, \bibinfo
  {author} {\bibfnamefont {S.}~\bibnamefont {B{\"a}{\ss}ler}}, \bibinfo
  {author} {\bibfnamefont {P.}~\bibnamefont {Sergelius}}, \bibinfo {author}
  {\bibfnamefont {R.}~\bibnamefont {H{\"u}hne}}, \bibinfo {author}
  {\bibfnamefont {B.}~\bibnamefont {Rellinghaus}}, \bibinfo {author}
  {\bibfnamefont {C.}~\bibnamefont {Shekhar}}, \bibinfo {author} {\bibfnamefont
  {V.}~\bibnamefont {S{\"u}{\ss}}}, \bibinfo {author} {\bibfnamefont
  {M.}~\bibnamefont {Schmidt}}, \bibinfo {author} {\bibfnamefont
  {C.}~\bibnamefont {Felser}}, \bibinfo {author} {\bibfnamefont
  {B.}~\bibnamefont {Yan}}, \ and\ \bibinfo {author} {\bibfnamefont
  {K.}~\bibnamefont {Nielsch}},\ }\href@noop {} {\enquote {\bibinfo {title}
  {Chiral magnetoresistance in the weyl semimetal {NbP}},}\ } (\bibinfo {year}
  {2016}),\ \bibinfo {note} {arXiv:1610.01413}\BibitemShut {NoStop}%
\bibitem [{\citenamefont {Gooth}\ \emph {et~al.}(2017)\citenamefont {Gooth},
  \citenamefont {Menges}, \citenamefont {S{\"u}{\ss}}, \citenamefont {Kumar},
  \citenamefont {Sun}, \citenamefont {Drechsler}, \citenamefont {Zierold},
  \citenamefont {Felser},\ and\ \citenamefont {Gotsmann}}]{Gooth:2017}%
  \BibitemOpen
  \bibfield  {author} {\bibinfo {author} {\bibfnamefont {J.}~\bibnamefont
  {Gooth}}, \bibinfo {author} {\bibfnamefont {F.}~\bibnamefont {Menges}},
  \bibinfo {author} {\bibfnamefont {C.~S.~V.}\ \bibnamefont {S{\"u}{\ss}}},
  \bibinfo {author} {\bibfnamefont {N.}~\bibnamefont {Kumar}}, \bibinfo
  {author} {\bibfnamefont {Y.}~\bibnamefont {Sun}}, \bibinfo {author}
  {\bibfnamefont {U.}~\bibnamefont {Drechsler}}, \bibinfo {author}
  {\bibfnamefont {R.}~\bibnamefont {Zierold}}, \bibinfo {author} {\bibfnamefont
  {C.}~\bibnamefont {Felser}}, \ and\ \bibinfo {author} {\bibfnamefont
  {B.}~\bibnamefont {Gotsmann}},\ }\href@noop {} {\enquote {\bibinfo {title}
  {Electrical and thermal transport at the planckian bound of dissipation in
  the hydrodynamic electron fluid of {WP$_2$}},}\ } (\bibinfo {year} {2017}),\
  \bibinfo {note} {arXiv:1706.05925}\BibitemShut {NoStop}%
\bibitem [{\citenamefont {Altland}\ and\ \citenamefont
  {Bagrets}(2016)}]{Altland:2016}%
  \BibitemOpen
  \bibfield  {author} {\bibinfo {author} {\bibfnamefont {A.}~\bibnamefont
  {Altland}}\ and\ \bibinfo {author} {\bibfnamefont {D.}~\bibnamefont
  {Bagrets}},\ }\href {\doibase 10.1103/PhysRevB.93.075113} {\bibfield
  {journal} {\bibinfo  {journal} {Phys. Rev. B}\ }\textbf {\bibinfo {volume}
  {93}},\ \bibinfo {pages} {075113} (\bibinfo {year} {2016})}\BibitemShut
  {NoStop}%
\bibitem [{\citenamefont {Louvet}\ \emph {et~al.}()\citenamefont {Louvet},
  \citenamefont {Houzet},\ and\ \citenamefont {Carpentier}}]{SupplMat}%
  \BibitemOpen
  \bibfield  {author} {\bibinfo {author} {\bibfnamefont {T.}~\bibnamefont
  {Louvet}}, \bibinfo {author} {\bibfnamefont {M.}~\bibnamefont {Houzet}}, \
  and\ \bibinfo {author} {\bibfnamefont {D.}~\bibnamefont {Carpentier}},\
  }\href@noop {} {\enquote {\bibinfo {title} {Supplementary material},}\
  }\BibitemShut {NoStop}%
\bibitem [{\citenamefont {Gorbar}\ \emph {et~al.}(2014)\citenamefont {Gorbar},
  \citenamefont {Miransky},\ and\ \citenamefont {Shovkovy}}]{Gorbar:2014}%
  \BibitemOpen
  \bibfield  {author} {\bibinfo {author} {\bibfnamefont {E.~V.}\ \bibnamefont
  {Gorbar}}, \bibinfo {author} {\bibfnamefont {V.~A.}\ \bibnamefont
  {Miransky}}, \ and\ \bibinfo {author} {\bibfnamefont {I.~A.}\ \bibnamefont
  {Shovkovy}},\ }\href@noop {} {\bibfield  {journal} {\bibinfo  {journal}
  {Phys. Rev. B}\ }\textbf {\bibinfo {volume} {89}},\ \bibinfo {pages} {085126}
  (\bibinfo {year} {2014})}\BibitemShut {NoStop}%
\bibitem [{\citenamefont {Ashby}\ and\ \citenamefont
  {Carbotte}(2014)}]{Ashby:2014}%
  \BibitemOpen
  \bibfield  {author} {\bibinfo {author} {\bibfnamefont {P.~E.}\ \bibnamefont
  {Ashby}}\ and\ \bibinfo {author} {\bibfnamefont {J.~P.}\ \bibnamefont
  {Carbotte}},\ }\href@noop {} {\bibfield  {journal} {\bibinfo  {journal} {Eur.
  Phys. J. B}\ }\textbf {\bibinfo {volume} {87}},\ \bibinfo {pages} {92}
  (\bibinfo {year} {2014})}\BibitemShut {NoStop}%
\bibitem [{Note1()}]{Note1}%
  \BibitemOpen
  \bibinfo {note} {Note that the Lorentz broadening of the Sharvin conductance,
  $g(\mu ,b=0)$, requires a high-energy cutoff below which the Hamiltonian
  \protect \textup {\hbox {\mathsurround \z@ \protect \normalfont
  (\ignorespaces \ref {eq:model}\unskip \@@italiccorr )}} is defined, as
  opposed to the $b$ dependent correction described by Eq.~\protect \textup
  {\hbox {\mathsurround \z@ \protect \normalfont (\ignorespaces \ref
  {eq:g-broaden}\unskip \@@italiccorr )}}.}\BibitemShut {Stop}%
\bibitem [{\citenamefont {Groth}\ \emph {et~al.}(2014)\citenamefont {Groth},
  \citenamefont {Wimmer}, \citenamefont {Akhmerov},\ and\ \citenamefont
  {Waintal}}]{Groth:2014}%
  \BibitemOpen
  \bibfield  {author} {\bibinfo {author} {\bibfnamefont {C.~W.}\ \bibnamefont
  {Groth}}, \bibinfo {author} {\bibfnamefont {M.}~\bibnamefont {Wimmer}},
  \bibinfo {author} {\bibfnamefont {A.~R.}\ \bibnamefont {Akhmerov}}, \ and\
  \bibinfo {author} {\bibfnamefont {X.}~\bibnamefont {Waintal}},\ }\href@noop
  {} {\bibfield  {journal} {\bibinfo  {journal} {New Journal of Physics}\
  }\textbf {\bibinfo {volume} {16}},\ \bibinfo {pages} {063065} (\bibinfo
  {year} {2014})}\BibitemShut {NoStop}%
\bibitem [{\citenamefont {Delplace}\ \emph {et~al.}(2012)\citenamefont
  {Delplace}, \citenamefont {Li},\ and\ \citenamefont
  {Carpentier}}]{Delplace:2012}%
  \BibitemOpen
  \bibfield  {author} {\bibinfo {author} {\bibfnamefont {P.}~\bibnamefont
  {Delplace}}, \bibinfo {author} {\bibfnamefont {J.}~\bibnamefont {Li}}, \ and\
  \bibinfo {author} {\bibfnamefont {D.}~\bibnamefont {Carpentier}},\
  }\href@noop {} {\bibfield  {journal} {\bibinfo  {journal} {EPL}\ }\textbf
  {\bibinfo {volume} {97}},\ \bibinfo {pages} {67004} (\bibinfo {year}
  {2012})}\BibitemShut {NoStop}%
\bibitem [{\citenamefont {Xiao}\ \emph {et~al.}(2010)\citenamefont {Xiao},
  \citenamefont {Chang},\ and\ \citenamefont {Niu}}]{Xiao:2010}%
  \BibitemOpen
  \bibfield  {author} {\bibinfo {author} {\bibfnamefont {D.}~\bibnamefont
  {Xiao}}, \bibinfo {author} {\bibfnamefont {M.-C.}\ \bibnamefont {Chang}}, \
  and\ \bibinfo {author} {\bibfnamefont {Q.}~\bibnamefont {Niu}},\ }\href
  {\doibase 10.1103/RevModPhys.82.1959} {\bibfield  {journal} {\bibinfo
  {journal} {Rev. Mod. Phys.}\ }\textbf {\bibinfo {volume} {82}},\ \bibinfo
  {pages} {1959} (\bibinfo {year} {2010})}\BibitemShut {NoStop}%
\bibitem [{\citenamefont {de~Jong}(1994)}]{Jong:1994}%
  \BibitemOpen
  \bibfield  {author} {\bibinfo {author} {\bibfnamefont {M.~J.~M.}\
  \bibnamefont {de~Jong}},\ }\href {\doibase 10.1103/PhysRevB.49.7778}
  {\bibfield  {journal} {\bibinfo  {journal} {Phys. Rev. B}\ }\textbf {\bibinfo
  {volume} {49}},\ \bibinfo {pages} {7778} (\bibinfo {year}
  {1994})}\BibitemShut {NoStop}%
\bibitem [{\citenamefont {Abrikosov}(1998)}]{Abrikosov:1998}%
  \BibitemOpen
  \bibfield  {author} {\bibinfo {author} {\bibfnamefont {A.~A.}\ \bibnamefont
  {Abrikosov}},\ }\href@noop {} {\bibfield  {journal} {\bibinfo  {journal}
  {Phys. Rev. B}\ }\textbf {\bibinfo {volume} {58}},\ \bibinfo {pages} {2788}
  (\bibinfo {year} {1998})}\BibitemShut {NoStop}%
\bibitem [{\citenamefont {Abrikosov}(2000)}]{Abrikosov:2000}%
  \BibitemOpen
  \bibfield  {author} {\bibinfo {author} {\bibfnamefont {A.~A.}\ \bibnamefont
  {Abrikosov}},\ }\href@noop {} {\bibfield  {journal} {\bibinfo  {journal}
  {Europhys. Lett.}\ }\textbf {\bibinfo {volume} {49}},\ \bibinfo {pages} {789}
  (\bibinfo {year} {2000})}\BibitemShut {NoStop}%
\bibitem [{\citenamefont {Lu}\ \emph {et~al.}(2015)\citenamefont {Lu},
  \citenamefont {Zhang},\ and\ \citenamefont {Shen}}]{Lu:2015}%
  \BibitemOpen
  \bibfield  {author} {\bibinfo {author} {\bibfnamefont {H.-Z.}\ \bibnamefont
  {Lu}}, \bibinfo {author} {\bibfnamefont {S.-B.}\ \bibnamefont {Zhang}}, \
  and\ \bibinfo {author} {\bibfnamefont {S.-Q.}\ \bibnamefont {Shen}},\
  }\href@noop {} {\bibfield  {journal} {\bibinfo  {journal} {Phys. Rev. B}\
  }\textbf {\bibinfo {volume} {92}},\ \bibinfo {pages} {045203} (\bibinfo
  {year} {2015})}\BibitemShut {NoStop}%
\end{thebibliography}

%

\end{document}